# Strong resistance nonlinearity and third harmonic generation in the unipolar resistance switching of NiO thin films

S. B. Lee,[1] S. C. Chae, [1] S. H. Chang, [1] J. S. Lee,[2] S. Park,[3] Y. Jo,[3] S. Seo,[4] B. Kahng,[2] and T. W. Noh[1, a)]

[1]ReCOE & FPRD, Department of Physics and Astronomy, Seoul National University, Seoul 151-747, Korea

[2]Department of Physics and Astronomy, Seoul National University, Seoul 151-747, Korea

[3]Quantum Materials Research Team, Korea Basic Science Institute, Daejeon 305-333, Korea

[4]Samsung Advanced Institute of Technology, Suwon, Gyeonggi 440-600, Korea

We investigated third harmonic generation in NiO thin films, which exhibit unipolar resistance switching behavior. We found that the low resistance states of the films were strongly nonlinear, with variations in the resistance $R$ as large as 60%. This strong nonlinear behavior was most likely caused by Joule heating of conducting

---

[a)] Author to whom correspondence should be addressed. E-mail: twnoh@snu.ac.kr

filaments inside the films. By carefully controlling the applied dc bias, we obtained several low resistance states, whose values of the third harmonic coefficient $B_{3f}$ were proportional to $R^{2+w}$ (with $w = 2.07\pm0.26$). This suggested that the resistance changes of the NiO films were accompanied by connectivity changes of the conducting filaments, as observed in classical percolating systems.

Unipolar resistance switching (RS) occurs when two reversible resistance states exist under one bias voltage polarity.[1] This intriguing phenomenon has been observed in numerous materials, including many binary oxides.[2] Recently, unipolar RS has attracted a great deal of interest, partly because of its technological potential for nonvolatile memory.[2,3] Although its physical mechanism is still a matter of debate, it is widely accepted that it occurs due to the formation and rupture of conducting filaments under an external bias voltage.[2,4–10]

The cluster structure properties of the conducting filaments, such as connectivity and topology, are likely to play a very important role in unipolar RS.[8] However, to date there have been no serious attempts to investigate these properties. By examining conducting spots on a sample surface, conductive atomic force microscopy can probe the appearance and disappearance of percolating conduction filaments, but not how they are connected inside the bulk sample.[8] It is very difficult to investigate the cluster structure of the filaments using other microscopic tools, such as transmission electron microscopy,[7] because the filaments form very complicated and winding channels inside a three-dimensional sample. Therefore, our understanding of the cluster structure of conducting filaments in real RS materials remains relatively poor.

Here, we present an investigation of the nonlinear electrical responses of NiO

thin films using third harmonic generation (THG) techniques. Our NiO films showed unipolar RS and very strong nonlinear electrical responses due to Joule heating. We compared THG signals during multi-level switching processes, and found that they scaled with resistance. These observations suggested that the RS process was accompanied by changes in connectivity among the conducting filaments. We also compared the critical exponents of the THG signals with those of classical percolating systems.

We grew polycrystalline NiO thin films on Pt/Ti/SiO$_2$/Si substrates using dc magnetron reactive sputtering. Using photolithographic techniques, we grew top electrodes of thickness 80 nm and area $30 \times 30$ $\mu$m$^2$. The detailed fabrication methods for NiO thin films and Pt top electrodes were described elsewhere.[1] We measured the dc current-voltage ($I_{dc}$-$V_{dc}$) curves using a semiconductor parameter analyzer (Agilent 4155C; Agilent Technologies, Santa Clara, CA).

Figure 1(a) shows the $I_{dc}$-$V_{dc}$ curve of a Pt/NiO/Pt capacitor that exhibited characteristic unipolar RS. The NiO thin film was highly insulating in the pristine state. When we applied a $V_{dc}$ of 3–5 V (not shown here), $I_{dc}$ increased suddenly, which could have caused a dielectric breakdown. To prevent damage, we kept the current in the NiO film below the compliance level $I_{comp}$. Immediately after this process, which we named

Forming, the film entered a low resistance state (LRS). When $V_{dc}$ was increased above 0.5 V, the film changed from the LRS to a high resistance state (HRS); we called this the RESET process. As we increased $V_{dc}$ again in the HRS, the film changed back into the LRS for $V_{dc}$ between 1.5 V and 3.0 V; we called this the SET process. For both Forming and SET processes, we used an $I_{comp}$ of 1 mA.

By carefully controlling the RESET voltage, we could observe multi-level RS (Fig. 1[b]). Just after Forming, our NiO film was in the LRS, denoted "1" in Fig. 1(b). An abrupt resistance change was observed as we increased $V_{dc}$ slowly to 0.49 V. Note that the film was not changed into the HRS, but into another level of LRS, indicated by "2" in the figure. By carefully controlling $V_{dc}$, we obtained four intermediate LRS levels. When $V_{dc}$ was increased above 0.56 V, the sample finally changed into the HRS.

The $I_{dc}$-$V_{dc}$ curves in the LRS showed strongly nonlinear variation with resistance. As shown in Fig. 1(a), the film deviated strongly from Ohmic behavior (dashed line), particularly near the RESET voltage. As shown in the inset of Fig. 1(a), the nonlinearity of the film resistance $R_{dc}$ was 60%. As shown in the inset of Fig. 1(b), such nonlinearity persisted at all levels of the LRS. It is likely that the increase in $R_{dc}$ was derived from the temperature increase of conducting filaments due to Joule heating.

To obtain further insight, we investigated the THG responses due to Joule heating.[11–14] Suppose that a current $i$ flows through a conducting filament with resistance $r$. Due to the Joule heating effects, $r$ increases by $\Delta r = r_o \beta \Delta T$, where $r_o$ and $\beta$ are the zero-biased value and temperature coefficient of resistance for the conducting filament, respectively. The filament's temperature $T$ rises due to Joule heating, so $\Delta T$ is proportional to $ri^2$. With $\Delta r \propto r_o ri^2$, the voltage drop $v$ across the filament is $v \approx r_o i + K r_o ri^3$, where $K$ is a proportionality coefficient. When an ac current with a frequency $f$, $i = i_{ac}\cos(2\pi ft)$, flows through the resistor, $v(t) = r_o i_{ac}\cos(2\pi ft) + v_{3f}\cos(6\pi ft)$ where $v_{3f} = (i_{ac}\Delta r)/2$. Hence, by measuring the $3f$ component, we could investigate the resistance nonlinear response due to Joule heating.

We performed THG experiments by placing a NiO film in one arm of a Wheatstone bridge circuit. We supplied an ac voltage $V_{ac}$ generated by a lock-in amplifier (SR830; Stanford Research Systems, Sunnyvale, CA). We used the phase-locked harmonic detection function of SR830 to measure the $3f$ voltage component $V_{3f}$ of the film. We could also obtain $R$ for the NiO film independently by balancing the Wheatstone bridge circuit. Using the measured $R$ values, we determined $I_{ac} = V_{ac}/R$. We obtained most $V_{3f}$ data at $f = 10$ Hz and at room temperature. We also performed THG measurements at other $f$ values (i.e., 1 Hz–2 kHz). For a given value of $V_{ac}$, we found that $V_{3f}$ was almost independent of $f$.[11,12]

Figure 2 shows the behavior of $V_{3f}$ for different levels of the LRS. As the level number of the LRS increased, $V_{3f}$ also increased significantly. For a given value of $I_{ac}$, e.g., 3 mA, the $V_{3f}$ value could vary by a factor of around 300. This variation was much larger than the $R$ variation, which was around 3.5, indicating that $V_{3f}$ was very sensitive to the nonlinear response of the NiO film. As shown by the solid lines in Fig. 2, $V_{3f}$ was proportional to $I_{ac}^3$ at all LRS levels. This was consistent with our earlier arguments based on THG due to Joule heating. For later use, we defined the third harmonic coefficient $B_{3f}$ as $V_{3f}/I_{ac}^3$.

Careful analyses of $B_{3f}$ can provide new insights into changes in the cluster structure parameters, such as connectivity, of the conducting filaments during the RESET process.[11–14] For an inhomogeneous medium, the current distribution should not be uniform. We assumed that the response of the NiO films could be represented by a network with randomly distributed conducting filaments. While $R$ measures the second moments of the current distribution, $B_{3f}$ probes the fourth moments[11–14]

$$\frac{B_{3f}}{R^2} \propto \frac{\sum i_\alpha^4}{\left(\sum i_\alpha^2\right)^2} \ , \tag{1}$$

where $i_\alpha$ is the current flowing through each conducting filament $\alpha$. If there is no change in the connectivity of conducting filaments, the right term in Eq. (1) remains constant and $B_{3f}$ is proportional to $R^2$. As shown in Fig. 3(a), for each LRS, $B_{3f} \propto R^2$.

This experimental result indicated that there was no change in the connectivity of conducting filaments while the LRS level was constant.

When the connectivity of the conducting filaments changes, the $R$ dependence of $B_{3f}$ varies. Traditionally, electrical breakdown in semicontinuous metal films has been treated using percolation theory.[11–15] The configuration change in the connectivity of conducting components changes the current distributions (i.e., in Eq. [1]), leading to $B_{3f}/R^2 \propto R^w$.[11–14] According to classical lattice percolation theory,[16] $0.82 < w < 1.05$. For two-dimensional semicontinuous metal films,[12,13] it has been shown reported that $1.2 < w < 2.0$.

Figure 3(b) shows the dependence of $B_{3f}$ on $R$ for a given $V_{ac}$ value. The open triangles and squares indicate the $B_{3f}$ values at $V_{ac} = 0.1$ V and 0.3 V, respectively, for the five levels of LRS in Fig. 1(b). To test the reliability of our data, we repeated the THG experiments with other NiO films. These results are indicated with the solid symbols. We found that, during the RESET process, $B_{3f}/R^2 \propto R^w$ with $w = 2.07\pm0.26$. The nonzero value of $w$ indicates that the connectivity of the conducting filaments changed during the RESET process. The $w$-value was similar to that of ion milled Au films ($w = 2.0\pm0.1$).[17]

We analyzed data at different $V_{ac}$ and $f$, and found that the $w$-value remained the same. We also performed THG measurements of the films at different temperatures. Although the values of $B_{3f}$ at 10 K became larger by a factor of 10 than those at room temperature, the $w$-value was almost the same. These observations indicated that the $w$-value did not depend on experimental variables, such as $V_{ac}$, $f$, or $T$. As in classical percolation problems, this critical exponent may provide a new way to study the cluster structure of the conducting filaments inside unipolar RS material systems. There have been no reports of theoretical studies of the $w$-value of unipolar RS materials. Further studies of this critical exponent are therefore necessary.

In summary, we found that the NiO thin films exhibited resistance switching behavior, with strong nonlinear electric responses. Using third harmonic generation, we demonstrated that the nonlinear resistance responses were most likely caused by Joule heating effects. We found that the third harmonic signals of multi-level low resistance states scaled with resistance with a critical exponent value, as in some percolating systems. The results presented here suggest that third harmonic generation could be an effective indicator of how conducting filaments are connected and disconnected in unipolar resistance switching materials.

This work was supported financially by the Creative Research Initiatives

(Functionally Integrated Oxide Heterostructure) of the Ministry of Science and Technology (MOST) and the Korean Science and Engineering Foundation (KOSEF). S.B.L. acknowledges financial support from a Seoul Science Scholarship.

**Figure captions**

FIG. 1. (Color online) (a) Characteristic dc current-voltage ($I_{dc}$-$V_{dc}$) curve of NiO films showing unipolar resistance switching. The LRS had a nonlinear $I_{dc}$-$V_{dc}$ curve that deviated from Ohmic behavior (dashed line). (b) Changes in the $I_{dc}$-$V_{dc}$ curves of the NiO film during multi-level switching. For (a) and (b), we set the compliance current $I_{comp}$ to 1 mA to avoid permanent dielectric breakdown. Insets show the strong nonlinear changes of dc resistance $R_{dc}$ ($\equiv V_{dc}/I_{dc}$) for the NiO film.

FIG. 2. (Color online) Third harmonic voltages $V_{3f}$ for the five levels of LRS shown in Fig. 1(b). At all levels, $V_{3f}$ was proportional to $I_{ac}^3$ at low $I_{ac}$ (solid lines). The relation $V_{3f} \propto I_{ac}^3$ suggested that the third harmonic signals could have been caused by Joule heating.

FIG. 3. (Color online) Dependence of the third harmonic coefficient $B_{3f}$ on $R$ for (a) each level of LRS and (b) a given value of $V_{ac}$. The open triangles and squares in (b) correspond to the data shown by the open symbols in (a). We repeated the same experiment with other NiO films and plotted the results using solid symbols. Note that $B_{3f} \propto R^{w+2}$ with (a) $w \approx 0$ and (b) $w = 2.07 \pm 0.26$.

The English in this document has been checked by at least two professional editors, both native speakers of English. For a certificate, see:

http://www.textcheck.com/cgi-bin/certificate.cgi?id=zJGpQ9

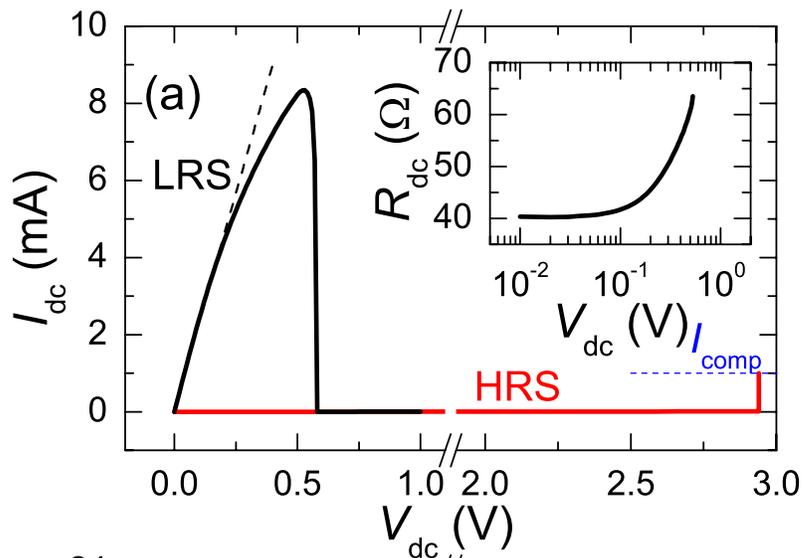

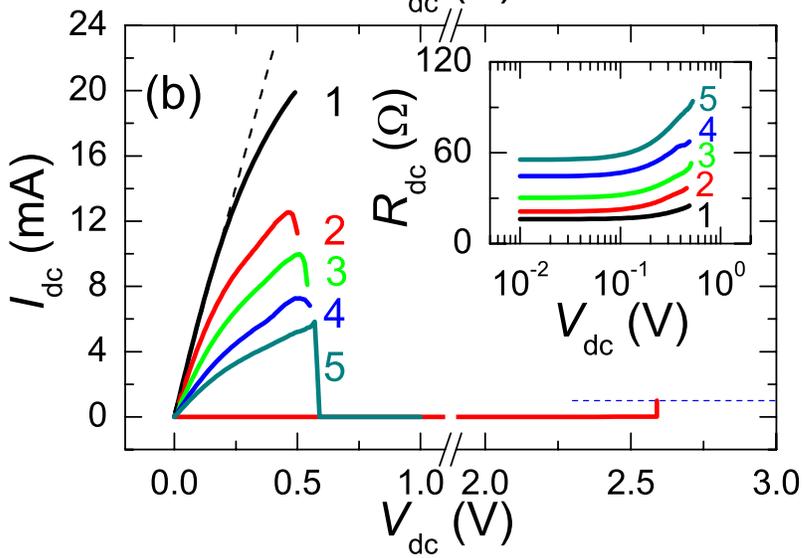

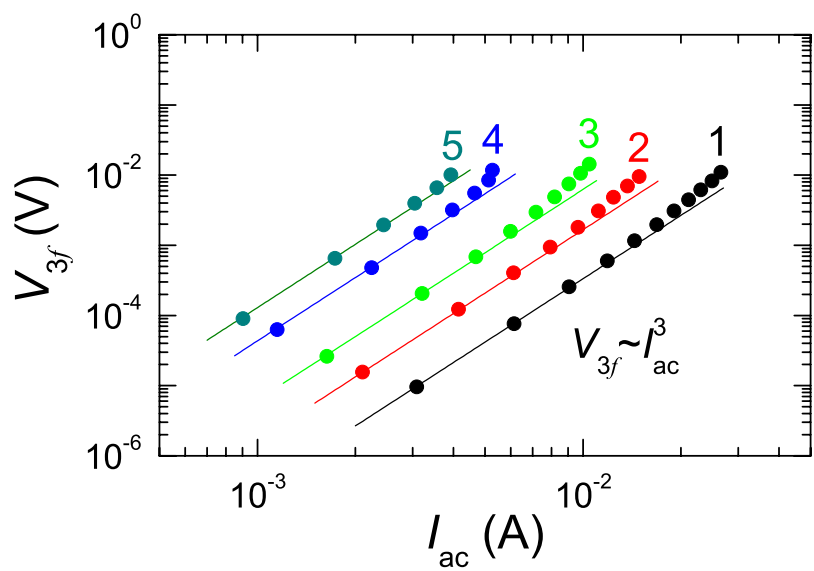

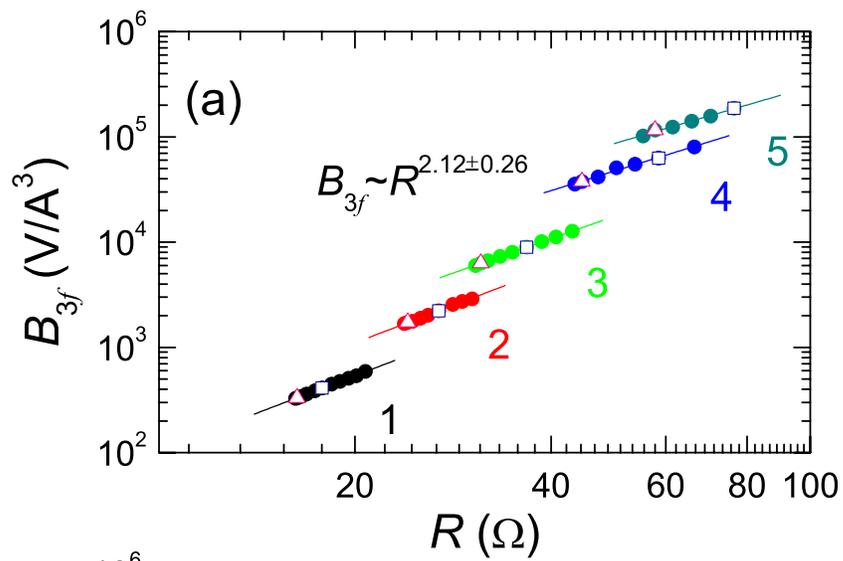

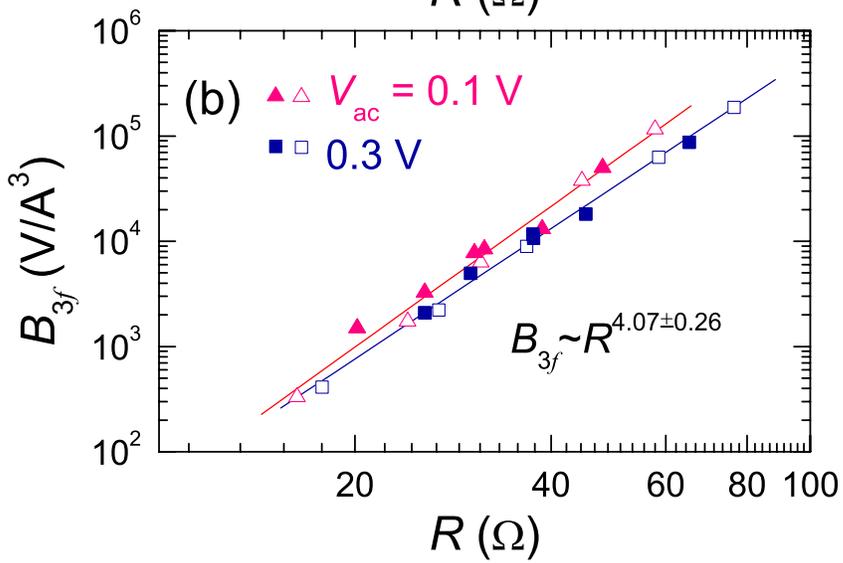